\documentstyle[hx37_latex,overcite]{article}

\input{psfig}
\def\be{\begin{equation}}
\def\ee{\end{equation}}
\def\bea{\begin{eqnarray}}
\def\eea{\end{eqnarray}}
\def\approxlt{\lower.2em\hbox{$\buildrel < \over \sim$}}
\def\approxgt{\lower.2em\hbox{$\buildrel > \over \sim$}}
\def\hi{{\rm H}\thinspace{\sc{i}}}
\def\hnought{\ifmmode H_0
    \else $H_0$\fi}
\def\kms{~{\rm km\ s}^{-1}}
\def\la{\ifmmode {{\rm Ly}\alpha}
        \else {Ly$\alpha$}\fi}
\def\msun{{$M_{\odot}$}}
\def\mkstar{M_{\rm K\ast}}
\def\civ{{\rm C}\thinspace{\sc{iv}}}
\def\oii{[{\rm O}\thinspace{\sc{ii}}]}
\def\ten#1{\ifmmode 10^{#1}
    \else $10^{#1}$\fi}

\begin{document}

\title{CONSTRAINTS ON GALAXY FORMATION FROM DEEP GALAXY REDSHIFT SURVEYS
AND QUASAR ABSORPTION LINE STUDIES}

\author{LENNOX L. COWIE}

\address{Institute for Astronomy, University of Hawaii, 2680 Woodlawn Dr.,\\
Honolulu, HI 96822, USA}


\maketitle\abstracts{ 
Magnitude-limited galaxy redshift surveys are now providing large
samples of galaxies to beyond $z=2$, while color-selected and
emission-line-selected samples are finding galaxies to $z=4.7$. A
broad picture is emerging of galaxy formation 
peaking in the $z=1$ range, which ties in with the
metallicity and density evolution seen in the quasar absorption
lines. We still have no direct information beyond $z=5$, but the
ionization of the IGM at this redshift argues for activity prior to
this time. The metallicities of around 0.01 solar which appear to be
relatively ubiquitous in quasar absorption lines beyond $z=2$, even
in very low column density clouds, could be a relict of this period.}

\section{Introduction}
A consistent picture of galaxy formation and evolution is emerging
from combining faint galaxy properties obtained from deep galaxy
surveys with what has been learned from quasar absorption line studies. 
At $z>2$ a large fraction of the baryon density of the universe
appears in quasar absorption line clouds stretching over neutral
hydrogen column densities from \ten{12} cm$^{-2}$ to \ten{22} cm$^{-2}$.  
This material has largely vanished by the present time, presumably
converting into stars.  Studies of the metallicity and ionization of the
baryonic material in absorbing clouds provide evidence for small amounts
of early
enrichment followed by a much more rapid rise below redshifts of about 2.  
At the same time, nearly complete spectroscopic surveys
are yielding information on the star-formation history
of galaxies, with evidence of substantial evolution in the properties
of individual galaxies  out to
redshifts of one, where there are many massive galaxies which
are dominated by the light of massive star formation. These also show
that the total star-formation rates in the universe were highest
near $z=0.5-1$  and were lower at higher redshifts
, although
sub-galactic or near-galactic sized star-forming galaxies exist at least
to redshift 4.7 [refs.~\citen{hu96a,hu96b}] and presumably extend beyond $z=5$. 

\section{Early Metals? }

\begin{figure}[h!]
\psfig{figure=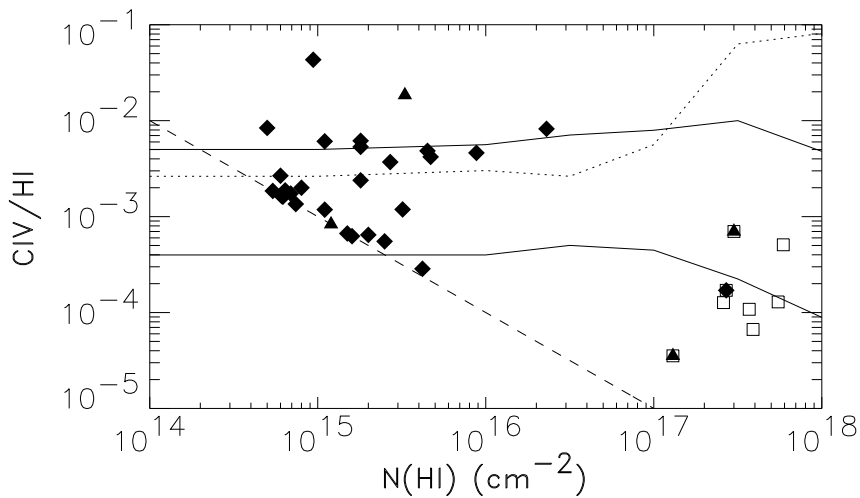,height=3.35in}
\caption{\protect{\civ} versus \protect{\hi} column density for all systems with
N(\protect{\hi}) $> 5 \times \ten{14}$ cm$^{-2}$\ at $3.135 < z < 3.60$\
toward Q1422+231(diamonds) and for N(\protect{\hi}) $\ge 1.5
\times 10^{15}~{\rm cm}^{-2}$\ at $z > 2.95$\ toward Q0014+813 (triangles).
Also shown (open squares) is \civ/\hi\ in all Lyman limit systems (open boxes) toward 
eight quasars \cite{so96}.  The dashed line shows the typical $2~\sigma$\
detection limit for \civ\ in Q1422+231.  The solid lines show model
calculations \cite{be86} of \civ/\hi\ for $\Gamma =
10^{-2.7}$\ (lower), $\Gamma = 10^{-1.7}$\ (upper) and the dotted line
the model for $\Gamma =
10^{-0.7}$, and a metallicity of $10^{-2}$\ solar, and illustrate that the
difference between the higher and lower column density systems is not solely a
radiative transfer effect but must arise from a higher ionization parameter or
higher metallicity in the weaker clouds.
\label{fig:metals}}
\end{figure}

One of the first results to emerge from the HIRES spectrograph
\cite{vogt} on the Keck telescope 
was the detection of many
weak \civ\ lines corresponding to relatively low column density
Lyman alpha
forest lines\cite{metals,tytler}.  This
provided confirmation of previous suggestions that there
might be fairly regular metal enrichment in such systems
\cite{meyer,tytler2}. Currently, as is summarised in Fig.~\ref{fig:metals},
we know that nearly all clouds with N(\hi) $>> \ten{15}$ cm$^{-2}$ and
a large fraction of those with N(\hi) $>> 3\times\ten{14}$ cm$^{-2}$ are 
detected in \civ\cite{metals2}. The issue of whether yet 
lower column density
clouds may be chemically unenriched remains open since
current sensitivity limits would not detect such clouds
in \civ\ at the expected column densities.

\vskip-0.6in

Remarkably the metallicity in the forest clouds of around
0.01 solar is similar to that seen in partial Lyman limit
system clouds with N(\hi) = $\ten{17}$ cm$^{-2}$ (Fig.~\ref{fig:metals}) and in the $z>2$ damped
\la\ systems with N(\hi)=$\ten{21}$ cm$^{-2}$ [ref.~\citen{lu}]. There
is also some evidence that the forest clouds show enhancement 
of the alpha process elements versus the Fe process elements\cite{metals2},
as may also be the case in the damped \la\ systems\cite{lu},
paralleling the abundances in the
low metallicity stars in the Galactic halo.

   This relatively ubiquitous enrichment clearly has powerful
implications for understanding the early stages of the 
heavy element formation. Essentially it requires that
along any line of sight through a forest cloud we must
see a relatively uniform (at least to order of magnitude)
enhancement, or alternatively expressed, that the covering
factor of metal-enriched portions of the clouds must be 
near unity. On smaller scales there may, of course, be lower
metallicity pockets or unenriched regions.

   We may, of course, be seeing directly the early stages of star formation
in the halos of forming galaxies but it is also possible,
perhaps even likely, that we may be seeing relicts left over
from subgalactic star formation in the dark ages at $z>5$.
If this is true, this metallicity and the fact that the
IGM is ionized at $z=5$ represent the two pieces of information
we have about this period. Even in the  $z=5$ era [ref.~\citen{rees}] it
is not easy to understand the uniformity since, based
on simple cooling arguments, substantial star formation
is only expected in near Galactic sized clumps.

\section{Galaxy Redshift Distributions }

\begin{figure}
\psfig{figure=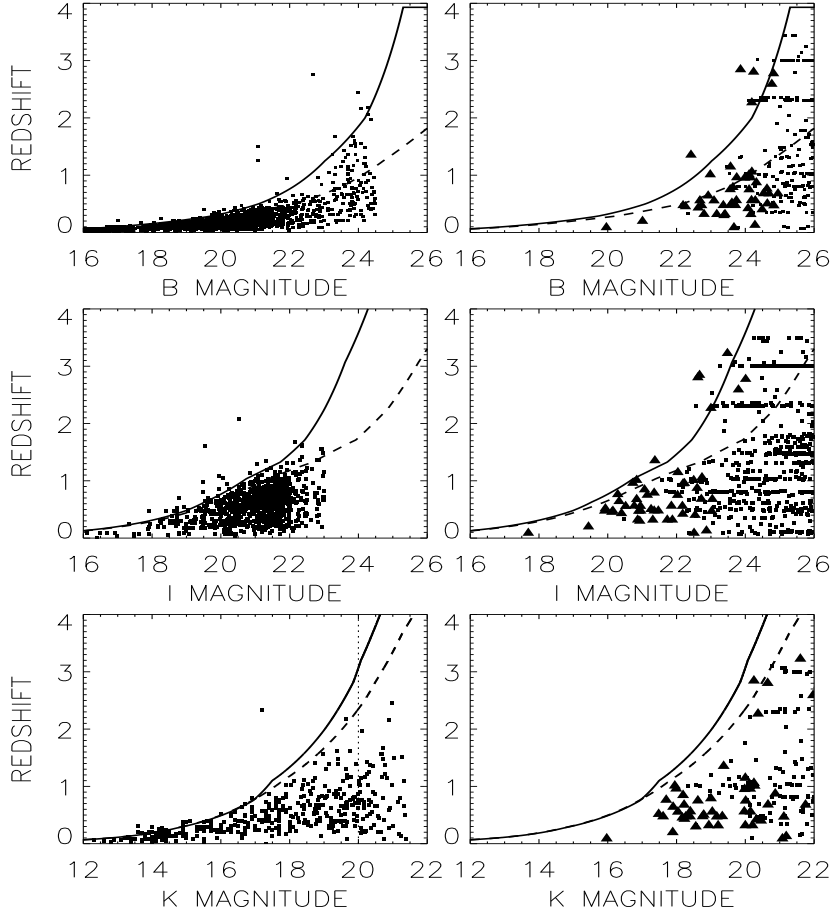,height=5.in,width=4.5in}
\caption{The left hand panels show the redshift magnitude diagrams
for near complete spectroscopic samples in the $K$, $I$(Kron Cousins)
and $B$ bands. The data is taken from the Autofib sample
of Ellis et al. \protect{\cite{ellis}} and the 
Hawaii survey \protect{\cite{cowie96}} for the $B$ sample, the CFRS
survey \protect{\cite{lilly}} and the Hawaii survey for the $I$ band
and the Glazebrook \protect{\cite{glazebrook}} sample and the Hawaii samples 
\protect{\cite{songaila,cowie96}} for
the $K$ band. For the $K$ band we have extended the sample beyond
the completeness limit of $K=20$ shown by the dotted line.
  The right hand panels show the spectroscopically identified
objects given in Cohen et al. \protect{\cite{cohen}} and Steidel et al
\protect{\cite{steidel}} in the Hubble Deep Field as triangles.  The remaining
objects are shown at color estimated redshifts using 
a six color estimator with the four Hubble Deep Field colors
and two IR colors ($J$ and $H+K$). This color estimator \protect{\cite{lilly}} agrees
well with all the measured spectroscopic redshifts.
The solid and dashed lines show an unevolving Im (solid) and
Sb (dashed) galaxy with $M_K=-25.8$, $M_I=-23.5$ and $M_B=-22$ for $\hnought=
50 \kms$/Mpc and $q_0=0.5$.
\label{fig:redshift_dis}}
\end{figure}

Currently, large and near-complete redshift samples exist
to $B=24.5$, $I=23$ and $K=20$ as is summarised in 
Fig.~\ref{fig:redshift_dis} in the
left hand panels. These surveys extend to about $z=2$ and show
that there is a very substantial evolution in the properties
of the galaxies even by redshift one. By $z=1$ there are many
large galaxies whose colors and emission line properties
show that they are dominated by massive star formation.
\cite{ellis,lilly,cowie96}

\begin{figure}
\psfig{figure=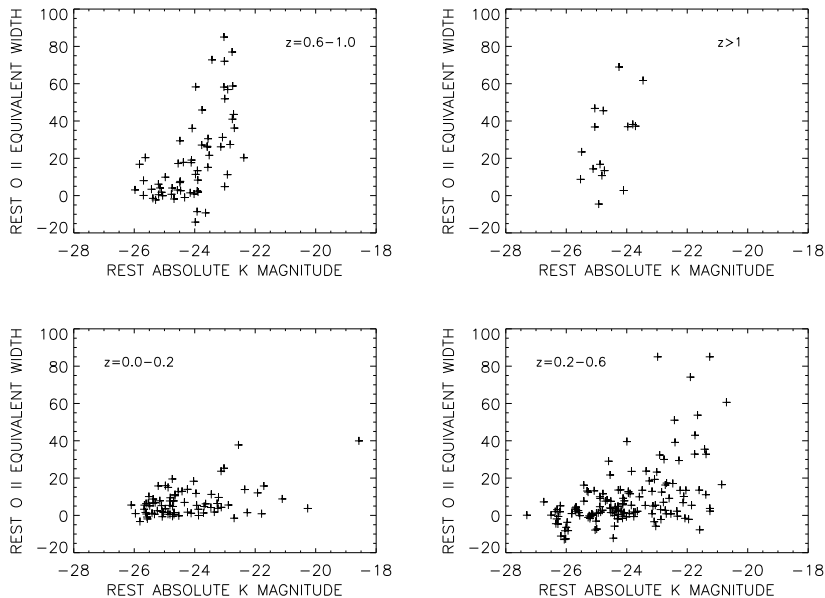,height=3.35in}
\caption{
Rest-frame \protect{\oii} equivalent width versus absolute rest $K$
magnitude and redshift for the $K < 20$\ sample.  In the lowest redshift
interval (lower left panel) very few galaxies have strong \protect{\oii} lines
or are undergoing rapid star formation (EW(\protect{\oii}) $\approxgt25$\AA).
At higher redshifts, progressively more massive galaxies are undergoing
rapid formation, until at $z>1$ the locus of rapidly forming galaxies
reaches a luminosity near $\mkstar\sim -25$. The absolute magnitudes
are calculated for $q_0=0.5$ and $\hnought=
50 \kms$/Mpc.
\label{fig:oii}}
\end{figure}

This may be most cleanly seen in the $K$-selected surveys
which do not have the bias against early type galaxies at
$z=1-4$ which is present even in red optical samples. Fig.~\ref{fig:oii} shows the rest equivalent width of
the \oii\ 3727 line versus absolute $K$ magnitude in 
various redshift slices. This figure shows an interesting
and (at least for  a simple minded CDM theorist) somewhat surprising
effect. Galaxies whose light is dominated by massive star formation
typically have \oii\ equivalent widths in excess of 25\AA\ [ref.~\citen{kenn}],
and we can see from Fig.~\ref{fig:oii} that there are very few such galaxies
at any $K$ luminosity at the present time. However at
$z=0.2$ there are small galaxies (the blue dwarfs) which
fall into the category of rapid star formers, and as we move to
higher redshift more and more massive galaxies appear to
be \oii\ luminous until at $z>1$ we see many near L* galaxies
($M_K*=-25.1$) in this category. There are very few super L*
galaxies at any redshift as can be seen from Fig.~\ref{fig:redshift_dis}.
Rather, galaxies seem to regulate at this value in their
earlier stages. An $M_I=-23.5$ galaxy has an $AB$ magnitude of
--23 and would be produced by a star formation rate of about
100 \msun/yr [ref.~\citen{cowie88}]. This could easily form a massive galaxy
if it persisted until $z=1$.

   In order to push this to higher redshifts we can construct
color-estimated redshifts from the Hubble Deep Field (HDF) as was
discussed by a number of groups at the conference and
in the literature\cite{yahil,gwyn}. The color estimates 
are not straightforward,
particularly at the faint end where galaxies generally
become very blue, and use of only the four Hubble colors
can make this very uncertain.
The estimates can be made somewhat more robust by extending
the wavelength coverage into the near IR and we have
used additional $J$ and $H+K$ colors to make a six color
estimate. These may be found at
{\bf http://www.ifa.hawaii.edu/$\sim$cowie/hdf.html}.
The right hand
panels of Fig.~\ref{fig:oii}. combine these color estimators
with spectroscopic data on the HDF to expand the magnitude
redshift relation. This may be compared with the 
gravitational lensing estimates discussed by Mellier
in this volume\cite{mellier}.

   With the exception of a scattering of AGN which lie above the curves,
we may see that even at faint magnitudes and
high redshifts galaxies remain at or below L*. 

\begin{figure}
\begin{center}
\
\psfig{figure=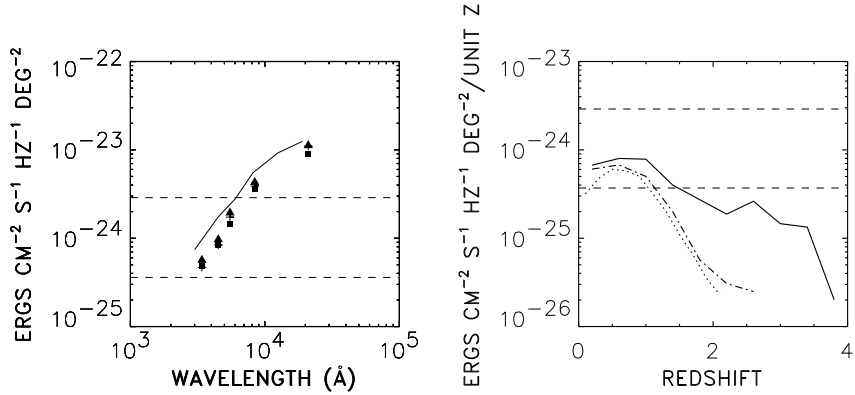,height=2.3in}
\caption{
The left panel shows the extragalactic background light (EBL) contribution
of the $I<26$ galaxy sample in the HDF as a function of wavelength
(solid line). Also shown is the same quantity computed for a $K<22$ sample
(triangles), an $I<24.5$ sample (crosses) and a $B<25.5$ sample (boxes)
in the Hawaii survey fields \cite{cowie_1}. The dashed lines show
the expected UV surface brightness required to form the local metal
density with the range reflecting the uncertainty in the metal
density estimates.\cite{cowie88,scl} The right hand panel shows
the evolution of the EBL per unit redshift 
at a wavelength of 3000(1+$z$)\AA\ versus
redshift in the $I<23$ spectroscopic
sample of the Hawaii fields (dotted line),
in the $I<23$ color estimated sample in the HDF (Dot-dash), and in 
the $I<26$ color estimated sample in the HDF (solid line).
\label{fig:surf}}
\end{center}
\end{figure}
\section{The history of galaxy formation}

The integrated extragalactic background light (EBL) from the galaxies in
the HDF is shown over the 3000\AA--20000\AA\ range in the
left panel of Fig.~\ref{fig:surf} where it is compared with the
integrated EBL in $B$-, $I$-, and $K$-selected samples from the
Hawaii Galaxy Survey fields.  The extra depth of the HDF does not
greatly increase the EBL, although the shape is slightly bluer,
since the number counts strongly converge in the red light density,
while in the blue they only logarithmically diverge.

The ultraviolet EBL is a direct measure of the metal density production
which may be compared with the local value of the metal density in the
universe, whose rather uncertain range is shown 
by the dashed lines\cite{cowie88,scl}. When sliced by redshift
it
therefore produces a direct history of the star formation in the universe.
This method has a considerable advantage over the method of constructing 
the UV light density  as a function of redhsift which was
discussed by a number of
speakers at the conference \cite{madau96}. The EBL technique measures the 
integrated production
of metals either in total or in a given redshift interval independent of
cosmology.  By contrast, the UV luminosity density measures the production
rate and is sensitive to the cosmology adopted. It must also be integrated
back to study the total production.

As has been known for some years there is enough ultraviolet light to
account for most of the current galaxies in relatively recent formation.
The spectroscopic or color-estimated redshift distributions allow us to expand
this result by slicing in the redshift direction.  In the right-hand
panel we show the EBL per unit redshift at a wavelength of 3000(1+$z$)\AA\
as a function of redshift.  The dotted line shows this for the $I<23$
sample in the Hawaii Survey Fields using full spectroscopic data, while
the dashed-dotted line shows the $I<23$ sample for the HDF using the
color-estimated redshifts.  The agreement is satisfactory.  The solid
line then shows the $I<26$ sample from the HDF.  As can be seen by 
comparing the  $I<23$ sample with the $I<26$ sample, the $I<23$ sample
maps the ultraviolet EBL well to just beyond $z=1$ so that the $I<26$
sample should be good to around $z=3$.  The fall-off in the $I<26$ curve is 
therefore real and most star formation has taken place at $z\approxlt1.5$.
As can be seen from the thin dashed lines, the total amount of star formation
agrees with the present metal density range.  At the present redshift the
larger area Hawaii Survey suggests that the star formation rate has begun
to turn down.

From Fig. 4 we can see that most of the star formation occurs after
$z=1.5$ though it is probably begining to die out by $z=0$. As we
have seen in section 3 below
$z=1$ most of this star formation is in relatively evolved galaxies.
\cite{cowie96} The redshift range for the formation matches that
in which the baryons in the quasar absorption line systems vanish
and is also broadly consistent with the general history of chemical
evolution and star formation in our own galaxy.\cite{cowie88,fall}

\section*{Acknowledgments}
This work was primarily supported by grants AR-06377.06-94A,GO-
5401.01-93A and GO-05922.0194A from STScI. I would like to
thank Richard Ellis for supplying the autofib data in machine
form and Esther Hu for assistance.
\section*{References}

%
%
%

\end{document}